\documentclass[aps,prl,twocolumn,superscriptaddress,floatfix]{revtex4}
\usepackage{amsbsy,amssymb,amsmath,bm}
\usepackage{graphicx,color,rotate}
\usepackage{units}
\usepackage{epstopdf}
\usepackage{floatrow}
\usepackage{multirow}
\usepackage{latexsym}

\usepackage{subfigure}
\usepackage{epsfig}
\usepackage{float}

\begin{document}

\title{Detailed study on the Fermi surfaces of the type-II Dirac semimetallic candidates $TM$Te$_{2}$ (where $TM$ = Pd, Pt)}

\author{W. Zheng}
\affiliation{National High Magnetic Field Laboratory, Florida State University, Tallahassee, Florida 32310, USA}
\affiliation{Department of Physics, Florida State University, Tallahassee-FL 32306, USA}
\author{R. Sch{\"o}nemann}\email{schoenemann@magnet.fsu.edu}
\affiliation{National High Magnetic Field Laboratory, Florida State University, Tallahassee, Florida 32310, USA}
\author{N. Aryal}
\affiliation{National High Magnetic Field Laboratory, Florida State University, Tallahassee, Florida 32310, USA}
\affiliation{Department of Physics, Florida State University, Tallahassee-FL 32306, USA}
\author{Q. Zhou}
\affiliation{National High Magnetic Field Laboratory, Florida State University, Tallahassee, Florida 32310, USA}
\affiliation{Department of Physics, Florida State University, Tallahassee-FL 32306, USA}
\author{D. Rhodes}\altaffiliation{Present address: Department of Mechanical Engineering, Columbia University, New York, NY, USA}
\affiliation{National High Magnetic Field Laboratory, Florida State University, Tallahassee, Florida 32310, USA}
\affiliation{Department of Physics, Florida State University, Tallahassee-FL 32306, USA}
\author{Y. -C. Chiu}
\affiliation{National High Magnetic Field Laboratory, Florida State University, Tallahassee, Florida 32310, USA}
\affiliation{Department of Physics, Florida State University, Tallahassee-FL 32306, USA}
\author{K. -W. Chen}
\affiliation{National High Magnetic Field Laboratory, Florida State University, Tallahassee, Florida 32310, USA}
\affiliation{Department of Physics, Florida State University, Tallahassee-FL 32306, USA}
\author{E. Kampert}
\affiliation{Dresden High Magnetic Field Laboratory (HLD-EMFL), Helmholtz-Zentrum Dresden-Rossendorf, 01328 Dresden, Germany}
\author{T. F{\"o}rster}
\affiliation{Dresden High Magnetic Field Laboratory (HLD-EMFL), Helmholtz-Zentrum Dresden-Rossendorf, 01328 Dresden, Germany}
\author{T. J. Martin}
\affiliation{The University of Texas at Dallas, Department of Chemistry and Biochemistry, Richardson, TX 75080 USA}
\author{G. T. McCandless}
\affiliation{The University of Texas at Dallas, Department of Chemistry and Biochemistry, Richardson, TX 75080 USA}
\author{J. Y. Chan}
\affiliation{The University of Texas at Dallas, Department of Chemistry and Biochemistry, Richardson, TX 75080 USA}
\author{E. Manousakis}
\affiliation{National High Magnetic Field Laboratory, Florida State University, Tallahassee, Florida 32310, USA}
\affiliation{Department of Physics, Florida State University, Tallahassee-FL 32306, USA}
\author{L. Balicas}\email{balicas@magnet.fsu.edu}
\affiliation{National High Magnetic Field Laboratory, Florida State University, Tallahassee, Florida 32310, USA}
\affiliation{Department of Physics, Florida State University, Tallahassee-FL 32306, USA}
\date{\today}

\begin{abstract}

We present a detailed quantum oscillatory study on the Dirac type-II semimetallic candidates PdTe$_{2}$ and PtTe$_{2}$ \emph{via} the temperature and the angular dependence of the de Haas-van Alphen (dHvA) and Shubnikov-de Haas (SdH) effects. In high quality single crystals of both compounds, i.e. displaying carrier mobilities between $10^3$ and $10^4$ cm$^2$/Vs,  we observed a large non-saturating magnetoresistivity (MR) which in PtTe$_2$ at a temperature $T = 1.3$ K, leads to an increase in the resistivity up to $5 \times 10^{4}$ \% under a magnetic field $\mu_0 H = 62$ T. These high mobilities correlate with their light effective masses in the range of 0.04 to 1 bare electron mass according to our measurements. For PdTe$_{2}$ the experimentally determined Fermi surface cross-sectional areas show an excellent agreement with those resulting from band-structure calculations. Surprisingly, this is not the case for PtTe$_{2}$ whose agreement between calculations and experiments is relatively poor even when electronic correlations are included in the calculations. Therefore, our study provides a strong support for the existence of a Dirac type-II node in PdTe$_2$ and probably also for PtTe$_2$. Band structure calculations indicate that the topologically non-trivial bands of PtTe$_2$ do not cross the Fermi-level ($\varepsilon_F$). In contrast, for PdTe$_2$ the Dirac type-II cone does intersect $\varepsilon_F$, although our calculations also indicate that the associated cyclotron orbit on the Fermi surface is located in a distinct $k_z$ plane with respect to the one of the Dirac type-II node. Therefore it should yield a trivial Berry-phase.
\end{abstract}

\maketitle

\section{Introduction}

In the last few years, solid state systems have emerged as promising candidates for searching quasiparticles having properties akin to particles originally predicted in high energy physics such as the Dirac, Weyl and Majorana fermions. Massless Dirac fermions were discovered in graphene \cite{novoselov_two-dimensional_2005} and in the various topological insulators \cite{zhang_topological_2009, young_dirac_2012, ando_topological_2013}. In these and in the so-called type-I Dirac semimetals \cite{Cd3As2}, cone shaped electron and hole bands meet at a single point in momentum space, i.e. the Dirac node. Unlike Dirac fermions that have been observed in particle physics, Weyl fermions have been discovered only recently as quasiparticles within certain semimetals, whose associated Weyl nodes emerge in pairs of opposite chirality, or topological charges, at linear touching points between electron and hole bands. Similar to type-I Dirac points, type-I Weyl points emerge when either inversion or time reversal symmetry is broken as in the TaAs family of compounds \cite{xu_discovery_2015, lv_experimental_2015, weng_weyl_2015}. Recently, it was predicted the existence of the so-called type-II Weyl and Dirac fermions which break Lorentz invariance and occur at the touching points between electron and hole-pockets within the energy momentum dispersion associated with tilted Weyl/Dirac cones in $k$-space \cite{soluyanov_type-ii_2015}. Candidates predicted to display Weyl/Dirac type-II electronic dispersions are the transition metal dichalcogenides like (W, Mo)Te$_{2}$ \citep{ali_large_2014, rhodes_bulk_2017} and the diphosphides (Mo, W)P$_{2}$ \cite{autes_robust_2016, schonemann_fermi_2017, kumar_extremely_2017}.

\begin{figure*}[htb]
\begin{center}
		\includegraphics[width = \linewidth]{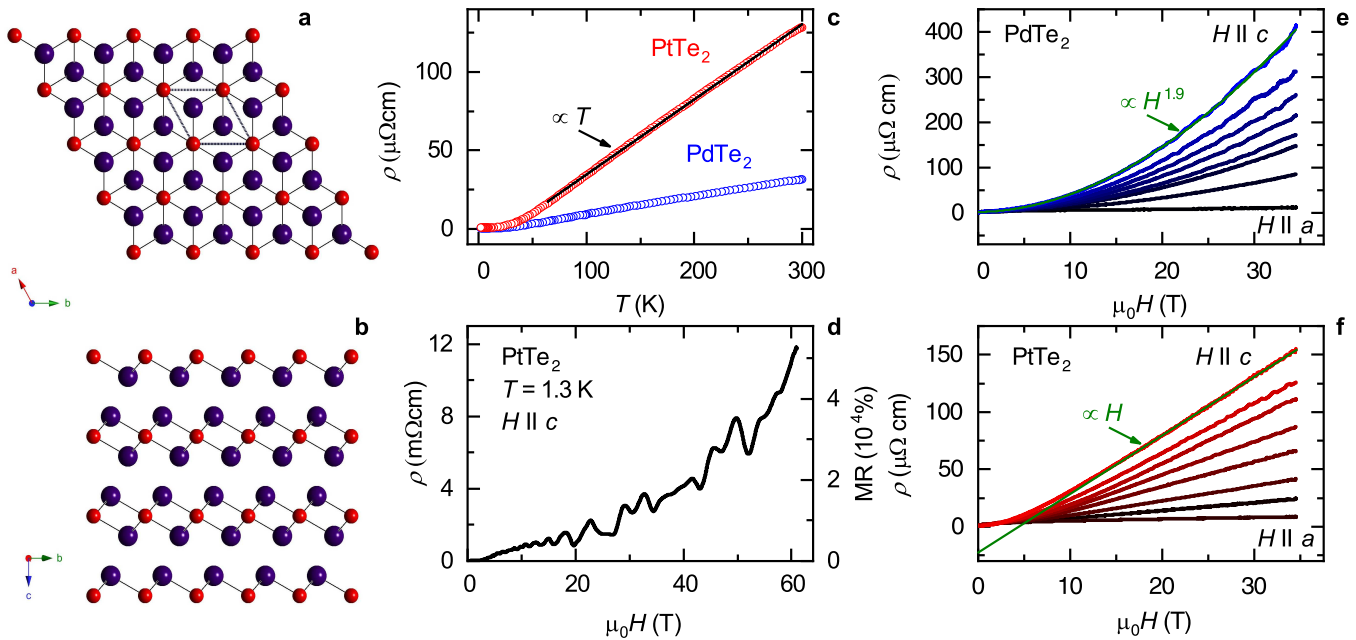}
		\caption{(a) and (b) Crystallographic structure of PdTe$_{2}$ (and of PtTe$_{2}$) \emph{via} the atomic arrangements observed within the \emph{ab} and \emph{bc}-planes, respectively. Here, the Pt/Pd and the Te atoms are symbolized by red and blue spheres, respectively. Unit cell is indicated by the grey dashed line in (a). (c) Resistivity $\rho$ as a function of the temperature for PtTe$_{2}$ and PdTe$_{2}$, red and blue makers respectively. Continuous black line is a linear fit, indicating $\rho \propto T$ for $T \gtrsim 30$ K. (d) $\rho$ as a function of the magnetic field $\mu_0H$ for fields up to $62$ T for a PtTe$_{2}$ single-crystal at $T = 1.3$ K and for $\mu_0 H\parallel c$-axis, as measured under pulsed magnetic fields. Oscillations in $\rho(\mu_0H)$ become markedly visible beyond $\simeq 10$ T due to the SdH-effect. (e) and (f) Magnetoresistivity curves for PdTe$_{2}$ and PtTe$_{2}$ in blue and in red at $T = 0.35$ K and for various angles between $H\parallel c$ and $H \parallel a$. Green lines represent fits of the high field data to a single power law $\rho(H) \propto H^{\alpha}$ yielding $\alpha = 1.9$ for PdTe$_{2}$ and $\alpha = 1$ for PtTe$_{2}$.}
	\label{fig:panel1}
\end{center}
\end{figure*}

The transition-metal dichalcogenides PdTe$_2$ and PtTe$_{2}$ crystalize in a layered CdI$_{2}$-type of structure within the trigonal space group $P\overline{3}m1$ \cite{revolinsky_layer_1963, furuseth_redetermined_1965, mangin_ptte2:_2008} which is shown in Figs. \ref{fig:panel1} (a) and (b). The crystal has inversion symmetry, hence the bands are Kramer's degenerate. PdTe$_{2}$ and PtTe$_{2}$ as well as PtSe$_{2}$ are predicted to be the first candidates for the realization of Dirac type-II fermions based on first principles calculations \cite{huang_type-ii_2016}. Recent publications claim to find experimental evidence for the existence of Dirac type-II points after mapping their electronic bandstructure \emph{via} ARPES measurements \cite{liu_identification_2015, yan_lorentz-violating_2017, noh_experimental_2017, zhang_experimental_2017} and after extracting the Berry-phase from magnetization measurements \cite{fei_nontrivial_2017}. Interestingly, in PdTe$_{2}$ Dirac type-I points are predicted to emerge upon application of hydrostatic pressure and coexist with the type-II points within a certain range of pressures \cite{xiao_manipulation_2017}.

\begin{figure*}
\begin{center}
		\includegraphics[width = \linewidth]{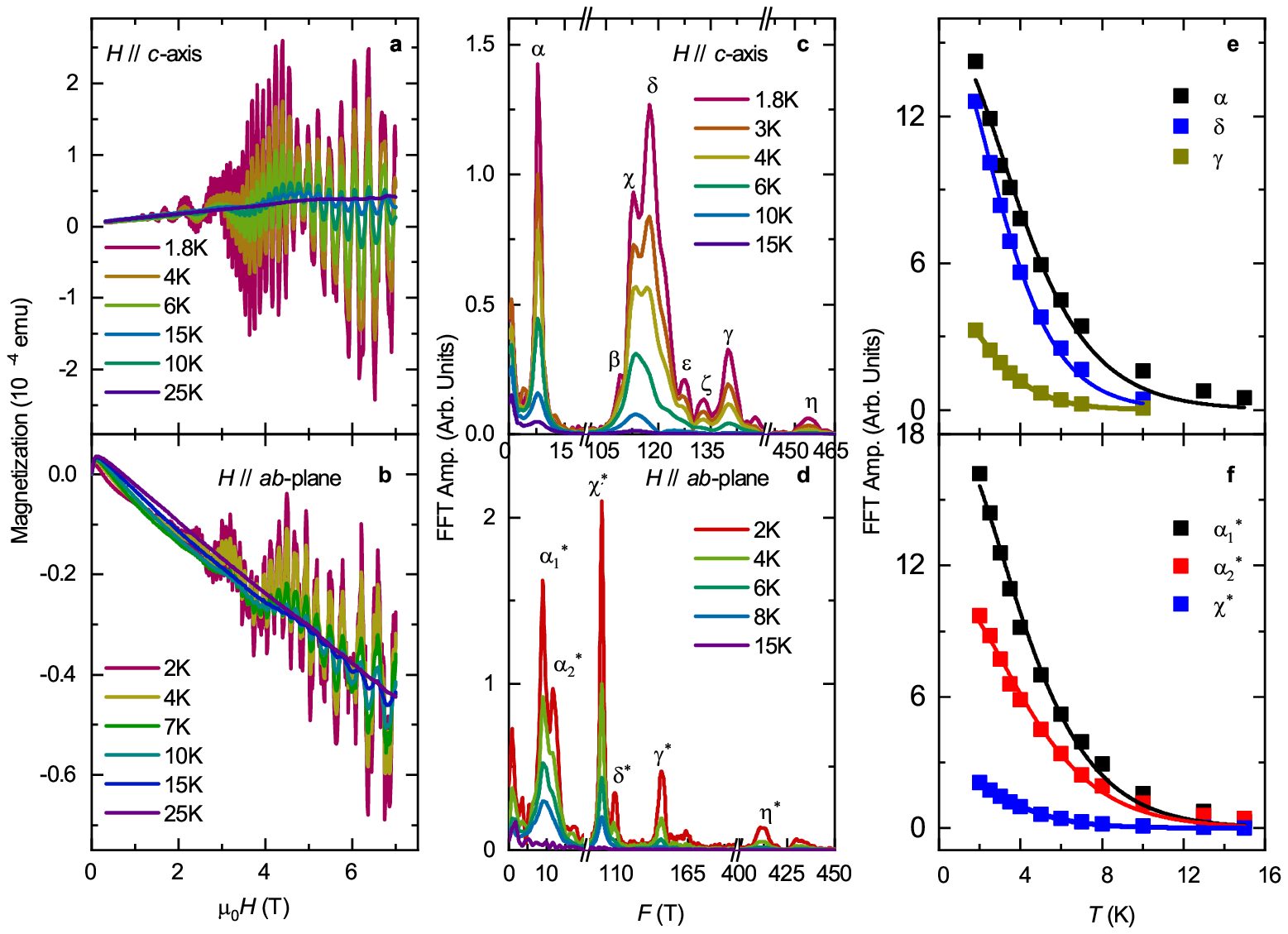}
		\caption{(a) and (b) Magnetization of PdTe$_{2}$ as a function of the magnetic field for two different field orientations, respectively and for several temperatures between $T = 1.8$ and 25 K. The corresponding Fast Fourier Transforms (FFTs) are shown in panels (c) and (d), respectively. Peaks corresponding to extremal cross sectional areas of the Fermi surface are labeled with Greek letters. (e) and (f) Amplitude of the main peaks observed in the FFT spectra as a function of $T$ where solid lines represent fittings of the experimental data by the temperature damping factor within the Lifshitz-Kosevitch formulism.}
	\label{fig:panel2}
\end{center}
\end{figure*}
PdTe$_{2}$ undergoes a superconducting transition below $T_c = 1.7\,\mathrm{K}$. In combination with topological non-trivial bands, this could make this compound a promising candidate for a topological non-trivial superconducting state, i.e. having an unconventional pairing symmetry. Its superconducting transition temperature $T_{\mathrm{c}}$ can be raised up to $4.65$ K \emph{via} Pd substitution with Au \cite{kudo_composition-induced_2016}. Magnetization and transport measurements indicate that PdTe$_{2}$ is a type-I superconductor \cite{leng_type-i_2017} thus supporting a recent report claiming conventional superconductivity for this compound \cite{das_conventional_2018}. Conventional $s$-wave superconductivity is also favored by recent heat capacity measurements in PdTe$_{2}$ \cite{amit_heat_2018} suggesting that its electronic structure might not in fact display a topological character.

Although the magnetoresistance and magnetization measurements of both PtTe$_2$ and PdTe$_2$ have been reported in the last year \cite{fei_nontrivial_2017, wang_hass-van_2016}, very detailed information about the Fermi surface topography extracted from angular dependence of bulk measurements like the de Haas-van Alphen (dHvA) and Shubnikov-de Haas (SdH) effects along with a comparison with Density Functional Theory (DFT) calculations are still missing. Here we present a complete Fermi surface (FS) study of both compounds through low field magnetization and high field electrical transport measurements in order to compare our experimental findings with first principle band-structure calculations. We find that DFT calculations describe the overall Fermi surface of PdTe$_2$ quite well, with some minor differences, but DFT provides a poor description for the
FS of PtTe$_2$. Our measurements confirm the existence of tilted Dirac type-II cones in PdTe$_2$, possibly also in PtTe$_2$. DFT also indicates for both compounds 
that the extremal-cross sectional areas of the Fermi surface detected by quantum oscillations ought to display a zero Berry-phase despite previous claims 
in favor of a topologically non-trivial cyclotron orbit.

\section{Experimental}

High quality single crystals of PdTe$_2$ were synthesized through a Te flux method: Pd (99.999 \%) and Te (99.999 \%) with an atomic ratio of 1:10 were sealed in an evacuated quartz ampule and subsequently heated to 800 $^{\circ}$C  and held at that temperature for 4 h. Then the ampule was slowly cooled to 525 $^{\circ}$C at a rate of 1 $^{\circ}$C/h. The excess Te was removed by centrifuging. The as harvested single-crystals were annealed for a few days at 520 $^{\circ}$C to improve the sample quality and remove residual excess Te. The synthesis of PtTe$_{2}$ followed qualitatively the same heating and annealing procedure. Pt (99.999 \%) and Te (99.999\%) in a ratio of 1:25 were heated up to 925 $^{\circ}$C, slowly cooled down to 600 $^{\circ}$C, and subsequently centrifuged. The synthesis yielded platelet like, easily cleavable single-crystals with the crystallographic $c$-axis oriented perpendicularly to the largest surface of the platelet. The phase purity of these crystals was confirmed by single-crystal X-ray diffraction (see, Supplemental Fig. S1 \cite{supplemental}) and energy-dispersive X-ray spectroscopy (EDX) measurements.

Conventional magneto-transport experiments on PdTe$_{2}$ and PtTe$_{2}$ single-crystals were performed in a Physical Property measurement System (PPMS) using a standard four terminal method under magnetic fields up to $\mu_0H = 9$ T and down to temperatures $T = 1.8$ K. High field magneto-transport experiments were performed in both a resistive Bitter magnet at the NHMFL in Tallahassee, under continuous fields up to $\mu_0H = 34.5$ T and temperatures down to $T \simeq 0.3$ K, and a pulsed magnet providing fields up to $\mu_0H = 62$ T with a pulse duration of 150 ms, at the Dresden High Magnetic Field Laboratory. Magnetization measurements under fields up to $\mu_0H = 7$ T were performed in a commercial superconducting quantum interference device (SQUID) magnetometer. Additional transport measurements were conducted in a $\mu_0 H= 18$ T superconducting magnet coupled to a dilution refrigerator. Magnetic torque measurements under pulsed magnetic fields were conducted with a piezoresistive microcantilever technique.

Electronic structure calculations were performed using the Quantum Espresso package \cite{giannozzi_quantum_2009} within the generalized gradient approximation (GGA) with the inclusion of spin orbit coupling (SOC). The structural parameters were taken from Ref. \cite{noauthor_materialsproject_nodate}. For the GGA+SOC calculations, fully relativistic norm conserving pseudo-potentials are generated using the optimized norm conserving Vanderbilt pseudopotentials as described in Ref. \cite{hamann_optimized_2013}. The 5$s$, 5$p$, 5$d$ and 6$s$ electrons for Pt, the $4s$, $4p$ and $4d$ electrons of Pd, and the 4$d$, 5$s$ and 5$p$ electrons for Te were treated as valence electrons. After rigorous convergence testing, the plane wave energy cutoff was taken to be 60 Ry and a $k$-point mesh of $11 \times 11 \times 8$ was used to sample the Brillouin Zone. The Fermi surfaces were generated using a $k$-point mesh of $51 \times51 \times 40$ and were visualized using the XCrysden software \cite{kokalj_xcrysdennew_1999}. The angular dependence was calculated using the Skeaf code \cite{rourke_numerical_2012}.

\section{Results}

Resistivity $\rho$ measurements as a function of the temperature on annealed PdTe$_{2}$ and PtTe$_{2}$ single-crystals are shown in Fig. \ref{fig:panel1} (c). All samples show metallic behavior over the entire temperature range, albeit with an anomalous linear dependence on $T$ above $T \simeq 30$ K. The large residual resistivity ratio, $RRR= \rho(300 \text{ K})/\rho(T \rightarrow 0\text{ K}) = 290$ for PdTe$_{2}$ and = 220 for PtTe$_{2}$ along with corresponding low residual resistivities $\rho_0$ are strong evidence for the high quality of these crystals. Here, $\rho(300 \text{ K})$ is the resistivity at 300 K and $\rho (T \rightarrow 0\text{ K})$ is the residual resistivity in the limit of zero temperatures as extracted from the resistivity data; $\rho(T)$  quickly saturates at the value of $\rho_0$ below $T = 30$ K. Notice that the best PdTe$_{2}$ crystals display  $\rho_{0} \simeq 0.1$ $\mu\Omega\mathrm{cm}$ while one obtains $\simeq 0.5$ $\mu\Omega\mathrm{cm}$ for PtTe$_{2}$. Notice also that neither compound display Fermi liquid behavior or $\rho(T) \propto T^2$, since the low temperature behavior corresponds just to a simple cross-over from a linear temperature dependence to saturation of the resistivity upon cooling yielding $\rho(T) \propto T^{\sim 3}$. The origin of the linear resistivity has been a matter of intense debate in strongly correlated systems such as the cuprates \cite{nigel} or the heavy Fermion compounds \cite{greg}, and frequently ascribed to the scattering of carriers by an unspecified bosonic mode. It is, therefore, surprising to observe such a dependence, over a decade in temperature, in a compound expected to be weakly correlated. Both compounds exhibit a large, non-saturating and anisotropic magnetoresistance (MR) as shown through Figs. \ref{fig:panel1} (d) to (f). At $T = 2$ K and under $\mu_{0}H = 9$ T, the MR reaches a few thousand per cent for fields applied along the crystallographic $c$-axis. Its behavior can be described as a combination of a linear and quadratic in field terms: $\rho(H) = \rho(\mu_{0}H = 0 \text{ T}) + A\mu_0H + B(\mu_0H)^{2}$, with positive $A$ and $B$ coefficients. Unlike compensated semimetals \cite{Rhodes}, PdTe$_{2}$ and PtTe$_{2}$ do not show the conventional quadratic in field dependence, nor can their $\rho(\mu_0H)$ be described by a single power law. Instead, one must include a linear in field component, as previously observed in Dirac systems upon approaching the quantum limit \cite{Cedomir}, to describe the magnetoresistive behavior observed under fields up to $\mu_0H = 9$ T.

To gain further insight into the magnetoresistive behavior of PtTe$_{2}$, we conducted Hall-effect measurements under fields up to $\mu_0H = 9$ T and temperatures between 2 and 300 K. The results are displayed in the Supplemental Fig. S2 \cite{supplemental}. We extracted the electron and the hole carrier densities $n_{\mathrm{e}}$, $n_{\mathrm{h}}$, and their respective mobilities $\mu_{\mathrm{e}}$, $\mu_{\mathrm{h}}$, from fittings of $\rho_{xx}$ and $\rho_{xy}$ to the two-band model. At $T=2$ K the fits yield $n_{\mathrm{e}}=8.8 \times 10^{20}\mathrm{cm^{-3}}$, $n_{\mathrm{h}}=10.2 \times 10^{20}\mathrm{cm^{-3}}$ and $\mu_{\mathrm{e}}=0.55 \times 10^{4}\mathrm{cm^{2}/Vs}$, $\mu_{\mathrm{h}}=0.36 \times 10^{4}\mathrm{cm^{2}/Vs}$. These carrier mobilities are just a factor of $\sim 2$ smaller than those of WTe$_2$ \cite{Joe} but one order of magnitude smaller when compared to those of $\gamma$-MoTe$_{2}$ \cite{rhodes_bulk_2017, zhou_hall_2016} which we tentatively attribute to a more effective carrier backscattering by impurities, although all of these systems display comparable residual resistivities. Interestingly, for PtTe$_{2}$ we found the difference between the densities of holes and electrons to be larger than 10 \%, particularly at low temperatures where the fits to the two-band model yield more accurate results. This indicates that charge carrier compensation is not the leading mechanism producing the large magnetoresistivity observed in this compound, as claimed to be the case for other semimetals like WTe$_2$ and MoTe$_{2}$ \cite{ali_large_2014, rhodes_bulk_2017}, PtBi$_{2}$ \cite{gao_extremely_2017} and W/MoP$_{2}$ \cite{schonemann_fermi_2017, kumar_extremely_2017}. Remarkably, transport measurements in PtTe$_{2}$ provide no evidence for saturation either, even under pulsed fields as high as $\mu_0H = 62$ T applied along its $c$-axis (Fig. \ref{fig:panel1}(d)). For PdTe$_{2}$ we observe no saturation in the magnetoresistivity under fields as high as 35 T, see Fig. \ref{fig:panel1}(e). Reliable Hall-effect data for this compound will be presented elsewhere.

\begin{figure}
\begin{center}
		\includegraphics[width = \linewidth]{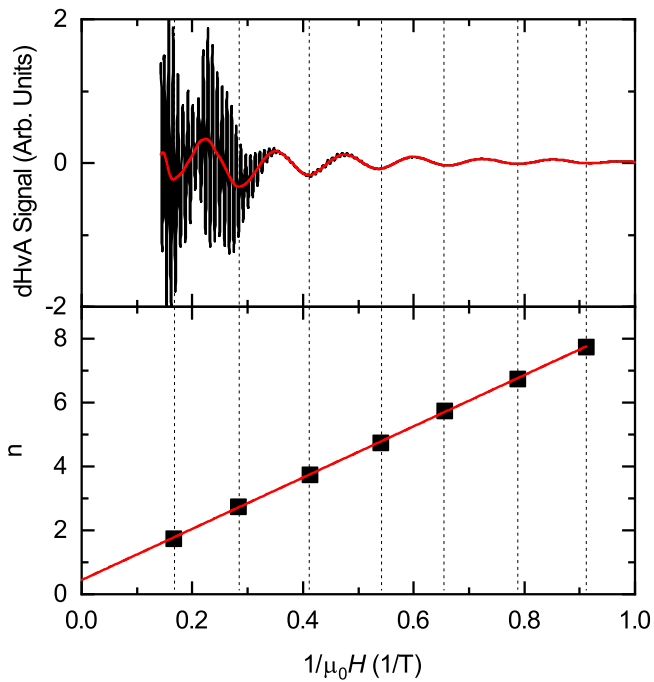}
		\caption{Top: The black line represents the raw oscillatory dHvA signal of PdTe$_{2}$ for $H\parallel c$. The oscillations of the lowest $\alpha$ frequency with $F_{\alpha}=8\,\mathrm{T}$ can be seperated by applying a low pass filter at $F=50$ T. The resulting oscillatory signal is shown as a solid red line. Bottom: Landau fan diagram of the $\alpha$ orbit.}
	\label{fig:panel4}
\end{center}
\end{figure}
\begin{table}
\caption{Effective masses of PdTe$_{2}$ and PtTe$_{2}$ from magnetization and transport measurements.}
\centering
\setlength{\tabcolsep}{6.5pt}
\renewcommand{\arraystretch}{1.1}
\begin{tabular}{c c c | c c c | c}
\hline\hline
\multicolumn{3}{c |}{PdTe$_{2}$, $H \parallel c$} & \multicolumn{3}{c |}{PdTe$_{2}$, $H \parallel a$} & \\
\hline
Orbit & $F(T)$ & $\mu/m_{0}$ & Orbit & $F(T)$ & $\mu/m_{0}$ & band\\
\hline
$\alpha$ & 8 & 0.045 & $\alpha_{1}^{*}$ & 9 & 0.051 & 3\\
$\beta$ & 109 & 0.039 & $\alpha_{2}^{*}$ & 12 & 0.048 & 2/4\\
$\chi$ & 113 & 0.045 & $\chi^{*}$ & 99 & 0.065 & 2/4\\
$\delta$ & 117 & 0.058 & $\delta^{*}$ & 106 & 0.041 & 2/4\\
$\epsilon$ & 127 & 0.056 & & & & 2/4\\
$\zeta$ & 133 & 0.060 & & & & 2/4\\
$\gamma$ & 140 & 0.069 & $\gamma^{*}$ & 145 & 0.077 & 2/4\\
$\eta$ & 455 & 0.075 & $\eta^{*}$ & 412 & 0.079 & 1\\
$\tau$ & 920 & 1.49 & & & & 3\\
$\phi$ & 2350 & 0.56 & & & & 3\\
$\kappa$ & 2675 & 0.69 & & & & 3\\
$\lambda$ & 3534 & 0.74 & & & & 2\\
$\nu$ & 5324 & 1.16 & & & & 3\\

\hline\hline
\multicolumn{3}{c |}{PtTe$_{2}$, $H \parallel c$} & \multicolumn{3}{c |}{PtTe$_{2}$, $H \parallel a$} & \\
\hline
$\alpha$ & 100 & 0.11 & $\alpha^{*}$ & 123 & 0.15 & 2\\
$\beta$ & 107 & 0.11 & & & & 2\\
$\gamma$ & 243 & 0.27 & $\gamma^{*}$ & 209 & 0.25 & 2\\
$\delta$ & 1703 & 1.6(2) & & & & 2\\
$\chi$ & 1971 & 1.6(2) & & & & 2\\
$\phi$ & 6068 & 3.6(8) & & & & 3\\
\hline
\end{tabular}
\label{table1}
\end{table}

Given the presence of Dirac type-II points within the electronic structure of both compounds one might expect to detect charge carriers characterized by topologically non-trivial Berry phases  \cite{liu_identification_2015, yan_lorentz-violating_2017, noh_experimental_2017, zhang_experimental_2017, fei_nontrivial_2017}. In fact, PdTe$_2$ was already claimed to be topologically non-trivial \cite{fei_nontrivial_2017} despite the DFT calculations (see Supplemental Information \cite{supplemental}) placing the Dirac type-II nodes deep below the Fermi level at -0.51 eV (-0.65 eV for PtTe$_2$). This is surprising since one might expect that the associated orbit might be located within a quadratically dispersing portion of the electronic band. But the authors of Ref. \onlinecite{fei_nontrivial_2017} extract a Berry phase $\phi_{B} \simeq \pi$ from the dHvA oscillations superimposed onto the magnetization, through the phase factor embedded within the Lifshitz-Kosevich (LK) formalism (see Ref. \cite{shoenberg_magnetic_1984} and also \cite{ando_topological_2013}). The LK formula describes the quantum oscillatory phenomena observed in density of states dependent physical variables such as the magnetization $M$ through:
\begin{equation}
\Delta M \propto -R_{T}R_{\mathrm{D}}R_S \sin\left(2\pi\left[\frac{F}{B}-\left(\frac{1}{2}-\phi\right)\right]\right),
\end{equation}

 here $R_{T} = \lambda T/\sinh(\lambda T)$ is the temperature damping factor, $R_{D} = \exp(-\lambda T_{\mathrm{D}})$ with $\lambda = 2\pi^{2}k_{\mathrm{B}}\mu / \hbar eB$ is the Dingle damping factor, and $R_S$ is the spin-damping factor \cite{shoenberg_magnetic_1984}. The phase factor $\phi = \phi_{\mathrm{B}}/2\pi - \delta$ contains the Berry phase $\phi_{\mathrm{B}}$ and a second phase-shift $\delta$ which takes values of 0 or $\pm 1/8$ (the sign depends upon the cross-sectional area, i.e. maxima or minima) for Fermi surfaces with two- and three-dimensional character, respectively. $T_{\mathrm{D}}$ is the sample dependent Dingle temperature and $\mu$ the effective cyclotron mass. To probe the topological character of these compounds, and check the claims of Ref. \onlinecite{fei_nontrivial_2017}, we measured the magnetization $M$ of PdTe$_{2}$ as a function of field $\mu_0 H$ for two field orientations, i.e. $H\parallel c$ and $H \parallel a$, as shown in Figs. \ref{fig:panel2} (a) and \ref{fig:panel2} (b), respectively. Similar measurements for PtTe$_{2}$ can be found in the Supplemental Fig. S3 \cite{supplemental}. The Fourier transforms (FFT) associated to the dHvA oscillations superimposed onto $M[(\mu_0H)^{-1}]$  are shown in Figs. \ref{fig:panel2}(c) and \ref{fig:panel2}(d), for both orientations. Several peaks are observable at frequencies $F$ between 8 and $500$ T for either field-orientation. In general, to extract the Berry phase associated with each individual orbit one would have to fit $\Delta M(\mu_0H)$ to a superposition of LK oscillatory terms, i.e. one for each cyclotron frequency. However, the large number of frequencies observed here (7 to 8) makes it nearly impossible to reliably extract the Berry-phase for individual orbits given that each LK component requires several input parameters (i.e. amplitude, frequency, phase, effective mass, quantum lifetime, etc.). Instead, we chose to apply a low pass filter around $F=50$ T to extract only the oscillations associated with the $\alpha$-orbit ($F_{\alpha} = 8$ T). The results are shown in the top panel of Fig. \ref{fig:panel4}. As discussed in Ref. \onlinecite{hu_evidence_2016}, the minima in the oscillatory dHvA signal can be assigned to Landau indices $n-1/4$ which produces the Landau fan diagram shown in the bottom panel of Fig. \ref{fig:panel4}. An extrapolation of the fan diagram to $1/(\mu_0H) \rightarrow 0$ leads to an intercept of 0.445, which is very close to the value 0.46  reported by Ref. \onlinecite{fei_nontrivial_2017}. From $(\phi_B/2\pi \pm \delta) = 0.445$ one obtains $\phi_B$ values of respectively $(2\pi \times 0.57)$ and  $(2 \pi \times 0.32)$, both $\neq (2 \pi \times 0.46 \sim \pi)$ reported by Ref. \onlinecite{fei_nontrivial_2017}. We cross-checked our procedure with data extracted for PtTe$_{2}$ and obtained the respective $\phi_{\mathrm{B}}$ values for the $\alpha$ and $\alpha^{*}$ orbits associated with topologically trivial electron pockets which are easy to separate from the other frequencies in the FFT spectra through the above filtering procedure. Both frequencies do yield topologically trivial Berry-phases.

\begin{figure}
\begin{center}
		\includegraphics[width = \linewidth]{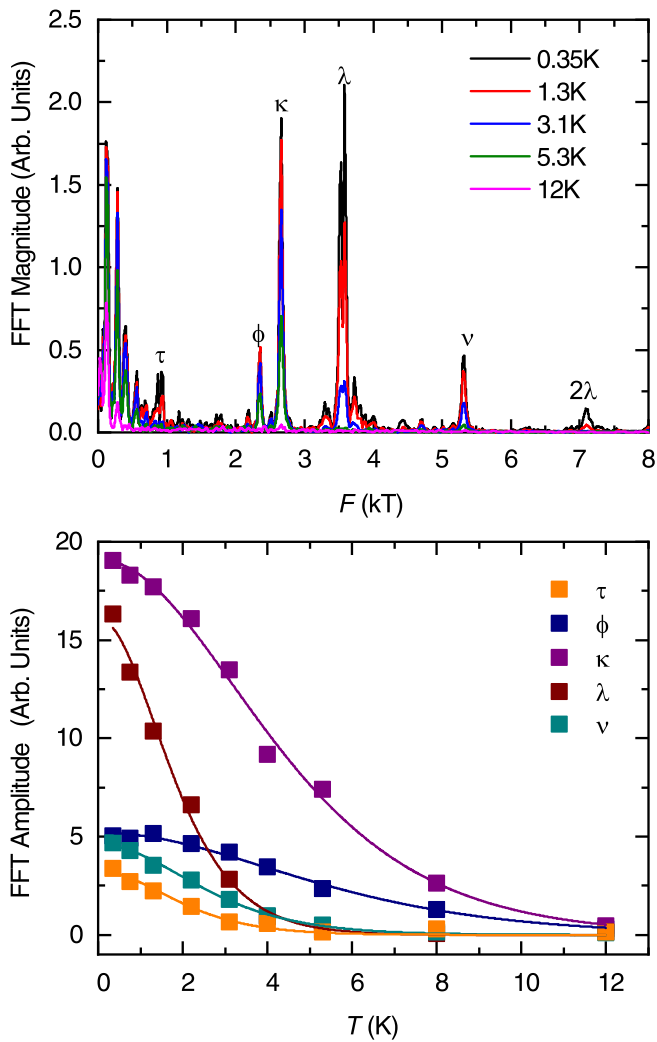}
		\caption{Fast Fourier transform (FFT) of the oscillatory component superimposed onto the $\rho(\mu_0 H)$ of PdTe$_{2}$ (see Fig. \ref{fig:panel1} (e)) for $\mu_0H\parallel c$-axis under temperatures $T$ ranging between 0.35 and 12 K. Peaks correspond to the extremal cross-sectional areas of the Fermi surface which are labeled with the Greek letters $\tau$, $\phi$, $\kappa$, $\lambda$ and $\nu$. Lower panel: amplitude of each peak/orbit observed in the FFT spectra as a function of $T$ where the solid lines are fits to the temperature damping factor $R_{T}$ in the LK formalism.}
	\label{fig:panel7}
\end{center}
\end{figure}
\begin{figure}
\begin{center}
		\includegraphics[width = \linewidth]{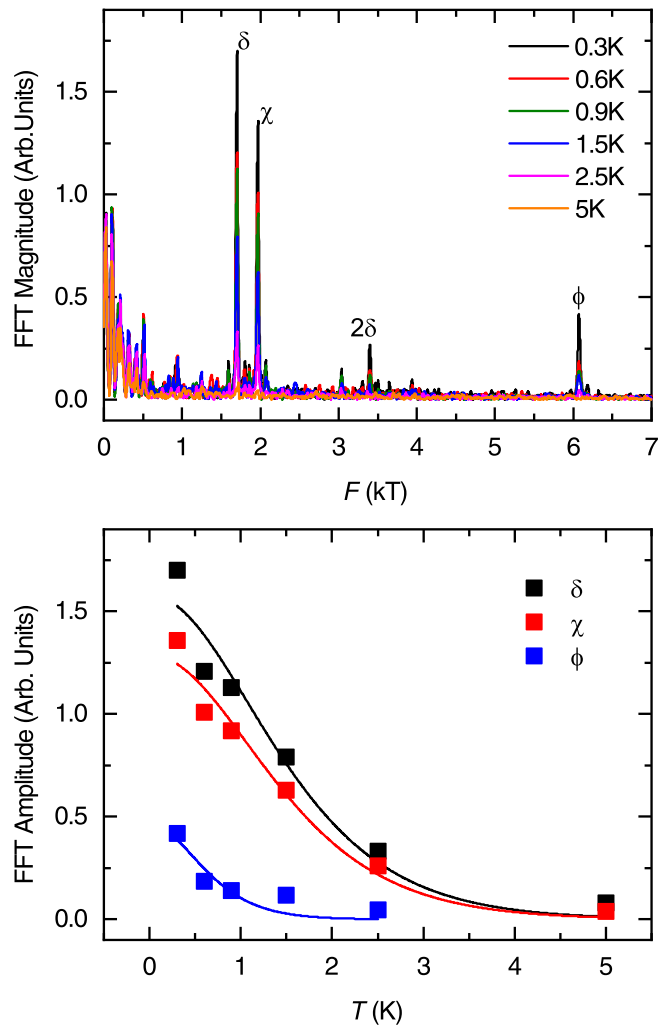}
		\caption{FFT of the oscillatory signal superimposed onto $\rho(\mu_0H)$ for PtTe$_{2}$ (see Fig. \ref{fig:panel1} (f)) for $H\parallel c$ at temperatures between 0.35 and 5 K. Greek letters $\delta$, $\chi$ and $\phi$ label the individual peaks associated to the extremal cross-sectional areas of the Fermi surface.  Lower panel: amplitude of each peak observed in the FFT spectra as a function of the temperature where solid lines represent fits to the temperature damping factor $R_{T}$ in the LK formalism.}
	\label{fig:panel8}
\end{center}
\end{figure}

The effective cyclotron masses for both PdTe$_{2}$ and PtTe$_{2}$ were determined from the amplitude of the dHvA and/or of the SdH oscillations as a function of the temperature. Lower dHvA frequencies become clearer in the magnetization measurements when $\mu_0 H \lesssim 7$ T  (see, Fig. \ref{fig:panel2} and also Supplemental Fig. S3). Depending on the orientation of the field, PdTe$_{2}$ and also PtTe$_{2}$, show either a paramagnetic or a diamagnetic background signal with superimposed quantum oscillations, indicating a rather anisotropic spin susceptibility. The effective masses $\mu_i$ can be extracted from the temperature dependence of the FFT amplitude of each individual peak by fitting the experimental data to the $R_{T}$ term in the LK formalism, as shown in Figs. \ref{fig:panel2}(e) and \ref{fig:panel2}(f). The effective masses associated to higher frequencies were obtained by analyzing the amplitude of the oscillations observed in transport experiments performed under high magnetic fields, see Figs. \ref{fig:panel7} and \ref{fig:panel8}. The extracted effective masses are summarized in Table \ref{table1}.

\begin{figure*}
\begin{center}
		\includegraphics[width = \linewidth]{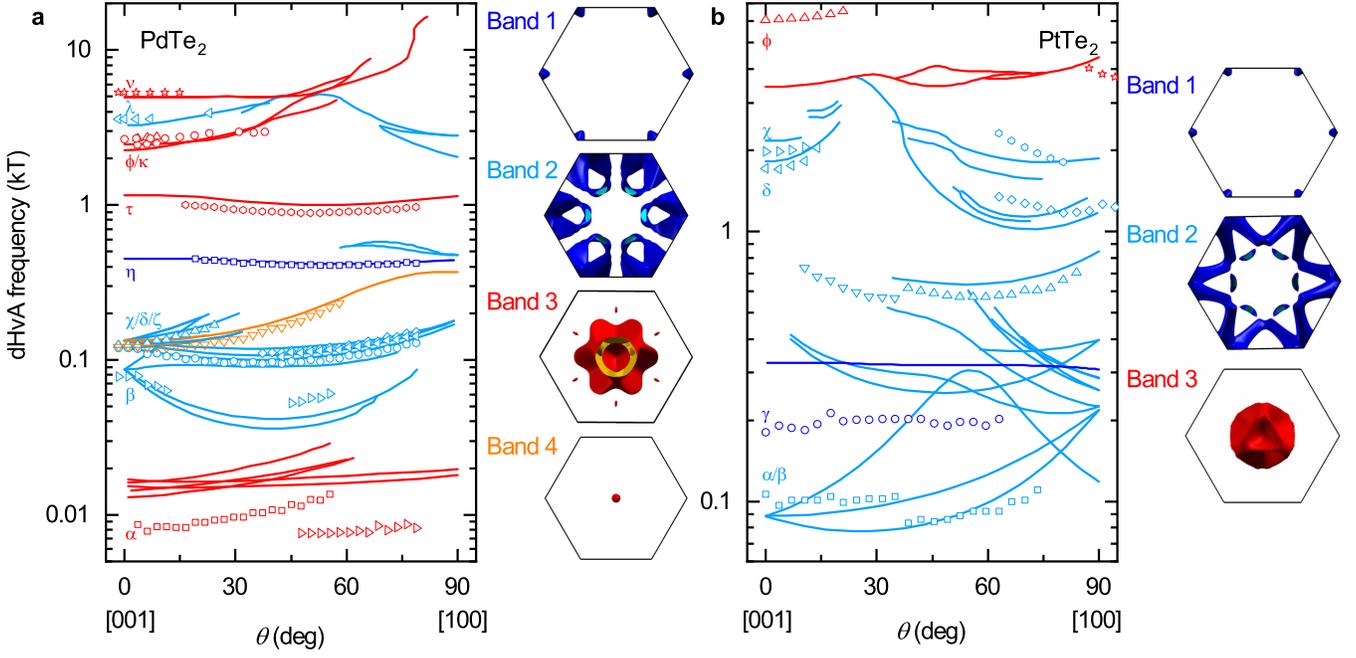}
		\caption{Quantum oscillatory frequencies as a function of the angle $\theta$ between $\mu_0 H$ and the \emph{c}-axis as extracted from PdTe$_{2}$ (a) and PtTe$_{2}$ (b) single-crystals and for fields rotating between $\mu_0H\parallel c$-axis to $\mu_0 H\parallel a$-axis. Symbols represent the experimental frequencies while solid lines depict their angular dependence according to the DFT calculations. For PdTe$_{2}$ bands 1 (in blue) and 2 (light blue) yield electron-like Fermi surface sheets, while bands 3 (red) and 4 (orange) lead to hole-pockets.  The same color scheme was applied to the bands of PtTe$_{2}$. The corresponding Fermi surfaces within the hexagonal first-Brillouin zone are displayed adjacent to each graph.}
	\label{fig:panel5}
\end{center}
\end{figure*}

From the magnetization of PdTe$_{2}$ and for both field orientations, we extract very light effective masses, i.e. between 0.039 and 0.075 $m_{0}$ for those orbits having frequencies below 500 T. Higher frequencies like $F_{\tau} = 920$ T and $F_{\phi} = 2350$ T yield effective masses of $\mu_{\tau} = 1.49$ $m_{0}$ and $\mu_{\phi} = 0.56$ $m_{0}$ and can be observed in the high field resistivity data for $H\parallel c$. They can be assigned to the larger hole pocket at the center of the Brillouin zone (band 3). PtTe$_{2}$ exhibits effective masses of 0.11 and 0.27 $m_{0}$ for the low frequencies $\alpha$, $\beta$ and $\gamma$, also with similar values for both field orientations. The $\delta$, $\chi$ and $\phi$ orbits, that can be assigned to the large electron and hole pockets (bands 2 and 3), lead to higher frequencies ranging between 1703 and 6068 T which display larger effective masses with values between 1.6 and 3.6 $m_{0}$.

The angular dependence of the frequencies extracted from the quantum oscillatory phenomena observed in PdTe$_{2}$ and in PtTe$_{2}$, as well as the angular dependence of the FS cross-sectional areas according to the DFT calculations, is shown in Fig. \ref{fig:panel5}, see also Supplemental Fig. S4 \cite{supplemental}. Most of the frequencies were extracted from the FFT of the oscillatory signal superimposed onto the resistivity measured at $T \simeq 25$ mK and under fields up to 18 T, see Supplemental Fig. S4. The agreement between the DFT calculations and the experimental data for PdTe$_{2}$ is remarkable, in particular for the $\eta$, $\chi$, and $\delta$ orbits resulting from the electron bands 1 and 2. The larger $\lambda$ orbit can only be observed for magnetic fields aligned nearly along the $c$-direction. This is also the case for the $\phi$, $\kappa$ and $\nu$ orbits, which can be assigned to the large open hole-pocket at the center of the Brillouin zone (band 3). For the lowest of the detected frequencies ($F \approx$ 8 T) labeled as the $\alpha$ orbit, which can be assigned to the small satellite pockets resulting from band 3, there is a sizeable mismatch between calculations and experiments. This disagreement is not surprising given that very small pocket areas like these tend to be very sensitive to small displacements in the Fermi level or in the precise position of the individual electronic bands which are within the energy resolution of the different DFT implementations.

\begin{figure*}[htb]
\begin{center}
\includegraphics[width= 17 cm]{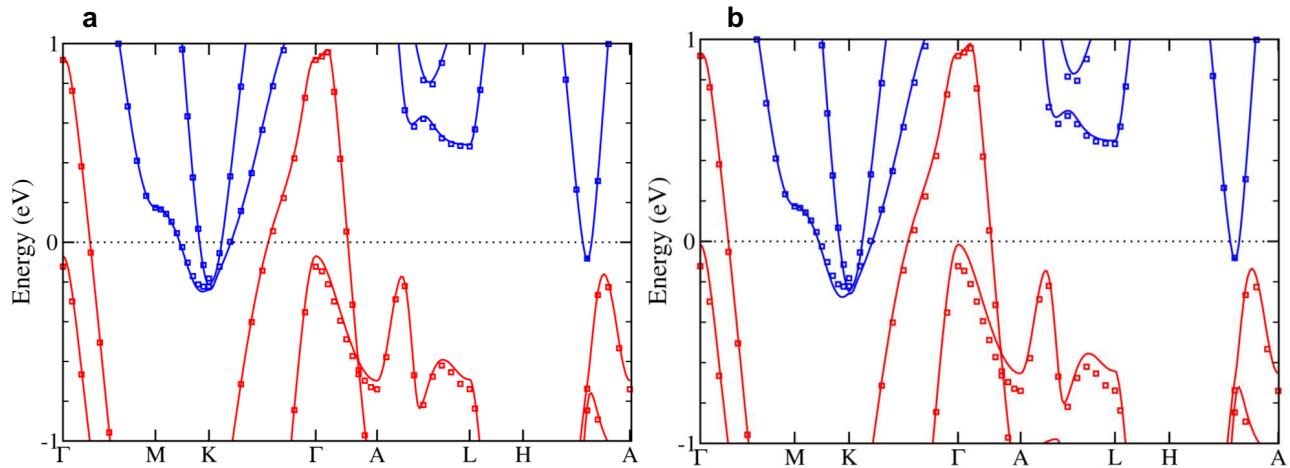}
\caption{(a) and (b) Band structure of PtTe$_2$ with spin-orbit coupling (SOC) when adding an on-site Hubbard $U$  on the
          Pt-$d$ site for the cases $U= 2$ and 4 eV, respectively. The open circles show the bands without $U$.}
   \label{fig:ptte2_bands_U}
\end{center}
\end{figure*}
In the case of PtTe$_{2}$ the experimental data cannot be as well described by the DFT calculations. Band 1, which creates nearly spherical hole surfaces at the edges of the Brillouin zone like in PdTe$_{2}$, can be assigned to the $\gamma$ frequency which has a nearly flat angular dependence, but with a value of $\sim 200$ T, instead of 325 T as predicted by DFT. In the region near 100 T we observe two frequencies, $\alpha$ and $\beta$, which nearly match those of the smaller pockets in band 2. Some of the frequencies above 500 T associated with the more complex shaped electron pockets of band 2 at the edge of the Brillouin zone can be matched with the $\delta$ and $\chi$ peaks observed when the field is nearly parallel to the $c$- or the $a$- axis. Given that the $\gamma$ and $\phi$ frequencies can be assigned to pockets in bands 1 and 3, one could increase the size of the hole sheets and decrease the size of the electron ones by lowering the Fermi energy which should improve the match between experiments and the calculations. However, we are not able to observe \emph{via} torque measurements a series of frequencies predicted within the range of 100 and 500 T, neither under static nor under higher pulsed fields. This might indicate that the shape of the Fermi surface is actually less complex than the predicted one. We confronted a similar situation when studying the Fermi surface of orthorhombic $\gamma$-MoTe$_2$ by observing a much simpler FS than the one extracted from angle resolved photoemission spectroscopy \cite{rhodes_bulk_2017}. Our unpublished calculations indicate that this discrepancy results from electronic correlations which motivated us to evaluate the role of correlations on the electronic structure and concomitant Fermi surface of PtTe$_2$.

\begin{figure*}[hb]
\begin{center}
\includegraphics[width= 11 cm]{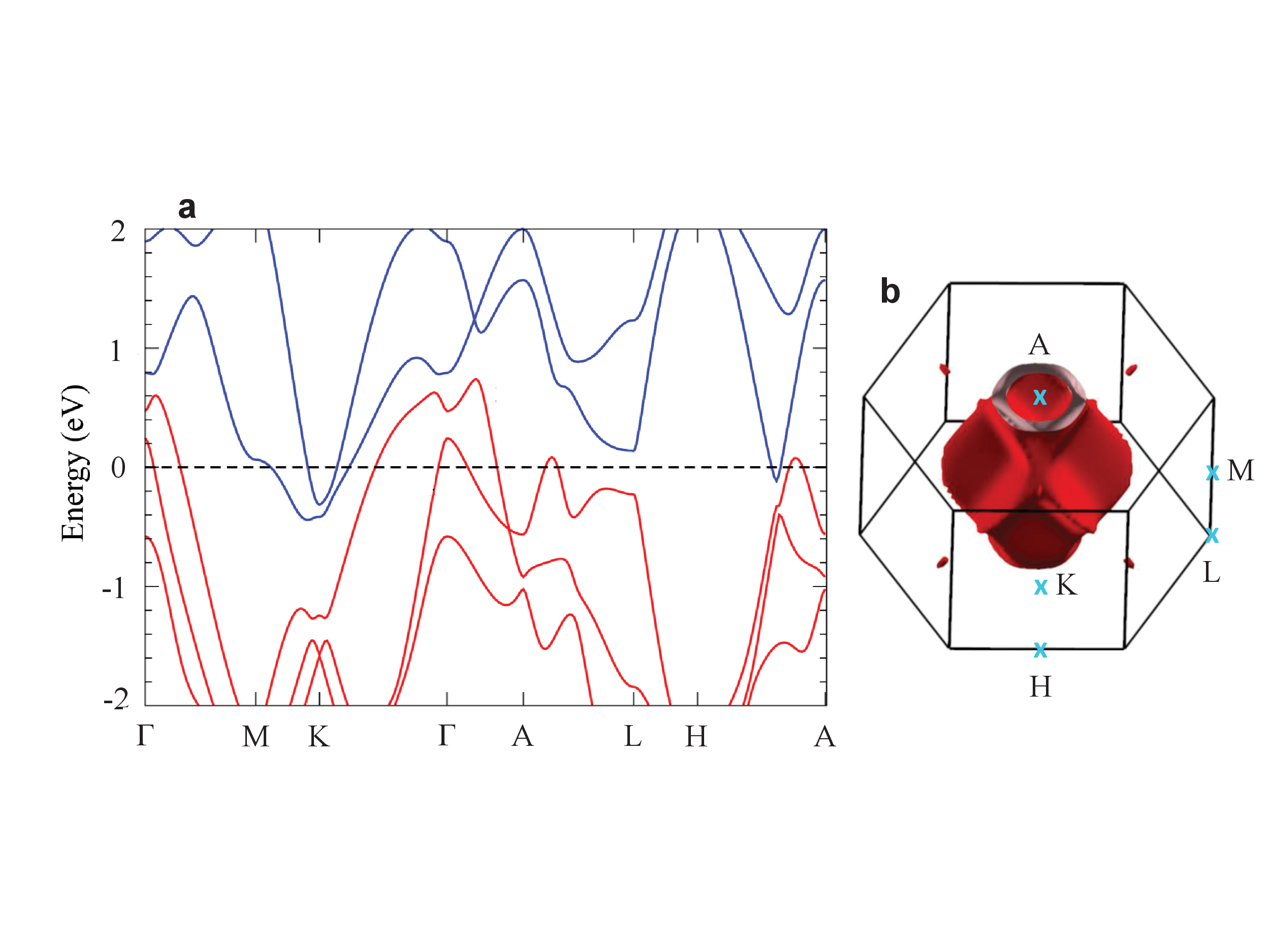}
\caption{(a) Band structure of PdTe$_2$. A transversal cut of the Dirac type-II cone is observed along the $\Gamma$-A direction with the node located at $\sim -0.5$ eV.  Notice that it disperses all the way up to the Fermi level. (b) Hole-like Fermi surfaces within the first Brillouin zone. A large hole sheet, responsible for the $\tau$ orbit, encloses the $\Gamma$-A direction along which the Dirac node is located. Clear blue crosses indicate the high symmetry points within the Brillouin zone.}
   \label{fig:pdte2_bands}
\end{center}
\end{figure*}
\section{Discussion}
\subsection {GGA+U calculations for PtTe$_2$}
As previously discussed, Fig.~\ref{fig:panel5} compares the angular dependence between the calculated and the experimentally measured Fermi surface cross-sectional areas
from quantum oscillation experiments. In contrast to PdTe$_2$, the agreement for PtTe$_2$ is far from excellent. Notice for example, that the orbit at around 6000 T is underestimated in the DFT calculations.

\begin{figure*}[htb]
    \begin{center}
             \includegraphics[width= 15 cm]{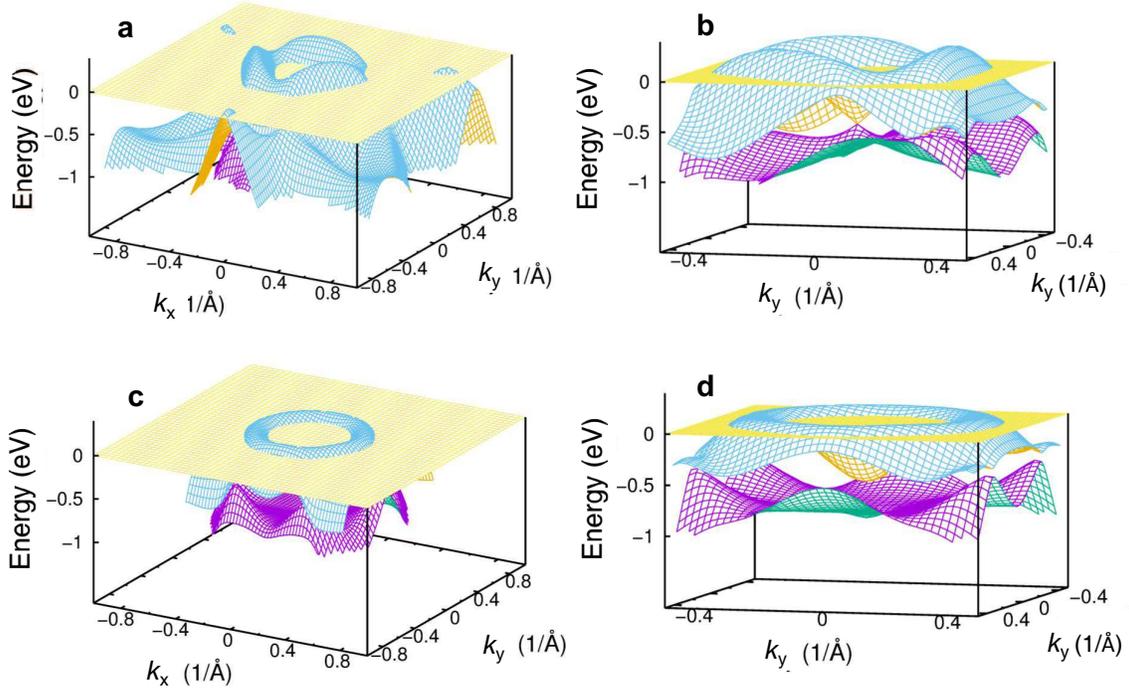}
        \caption{Two-dimensional band structure of PdTe$_2$ within the $k_x-k_y$ plane for different values of $k_z$. (a) and (b) show the bands at a value of
$k_z$ where the Dirac point occurs (i.e., at $k_z=k_z^{\text{DP}}$) whereas (c) and (d) show the bands for $k_z$ corresponding to the extremum orbit, i.e., the $\tau$ orbit, close to the Dirac point ($k_z=0.5 \times 2\pi/c$). Notice that the small islands in (a) have an area of $\approx 20$ T. These islands appear for $k_z$ values in the close vicinity of the $k_z^{\text{DP}}$ only. }
    \label{fig:bands_kx_ky}
  \end{center}
\end{figure*}
This orbit results from the large hole band centered around the $\Gamma$-point. Within the  GGA+SOC+U framework we have evaluated whether the reason for this discrepancy, in the case of the PtTe$_2$ compound, results from electronic correlations. Therefore, we used 2 eV and 4 eV for the value of an on-site Hubbard $U$ on the Pt-$d$ orbital. But as seen in Figs.~\ref{fig:ptte2_bands_U}(a) and \ref{fig:ptte2_bands_U}(b), respectively, for this range of U, we do not see any significant difference between band structures (with and without correlations) around the Fermi level. Therefore, the origin of this discrepancy between the calculated orbits and the experimentally measured ones remains to be resolved. Notice from Fig.~\ref{fig:ptte2_bands_U} that the bands crossing at $\sim -0.5$ eV, thus forming the Dirac type-II cone (located along the line from $\Gamma$ to A), do not disperse all the way up to $\varepsilon_F$. Therefore, the entirety of the Dirac cone does not intersect the Fermi level implying that PtTe$_2$ should, according to DFT, \emph{not} display topologically protected charge carriers. Subsequently, we discuss the case of PdTe$_2$ whose Dirac type-II cone does intersect $\varepsilon_F$, see Fig.~\ref{fig:pdte2_bands}.
\subsection {Dirac point and Berry phase in PdTe$_2$}
Our quantum oscillatory study does concede the possibility of a Berry phase being relatively close to $\pi$, but only for the orbit of size $\sim 10$ T and for magnetic fields parallel to the $c$-axis. From its angular dependence we identified it with the small ellipsoid-like orbit of size around $20$ T, according
to our DFT calculations (see Fig.~\ref{fig:pdte2_bands}(a)) which are seen in our Fermi surface calculations in Fig.~\ref{fig:pdte2_bands}(b) as the 4 small ellipsoid-like pockets.
In addition, the two-dimensional band surfaces near this pocket are shown by the small ``islands'' in Fig.~\ref{fig:bands_kx_ky}.
However, as we discuss below there is no reason to expect a non-trivial topology in this case. Furthermore, the extraction of Berry phase by quantum oscillation experiments
in the case of systems which possess both time-reversal-symmetry and inversion symmetry is plagued by the issues discussed
in Ref.~\onlinecite{kuan_wen}. As has been previously discussed, there is a type-II Dirac point (DP) which is shown in Fig.~\ref{fig:pdte2_bands}
along the $\Gamma-A$ line at about $0.5$ eV below the Fermi surface. The $k_z$ value corresponding to the position of this DP is found to be $0.4 \times 2\pi/c$
which we will be referred  as $k_z^{\text{DP}}$.


Since the position of the DP is about $-0.5$ eV from the Fermi level ($\varepsilon_{F}$), the detection of a Berry phase of $\sim \pi$ in other
reports \cite{fei_nontrivial_2017} is surprising and hence deserves further investigation. In Fig.~\ref{fig:bands_kx_ky} we present
the two-dimensional bands in the $k_x-k_y$ plane for two values of $k_z$.
First, in Fig.~\ref{fig:bands_kx_ky}(a) (Fig.~\ref{fig:bands_kx_ky}(b) is a zoomed-in version), the $k_z$ value is fixed at $k_z^{\text{DP}}$
and shows a type-I Dirac point at $(0,0)$. Fig.~\ref{fig:bands_kx_ky}(c) (Fig.~\ref{fig:bands_kx_ky}(d) is a zoomed-in version) presents the two-dimensional band structure for $k_z=0.5 \times 2\pi/c$, a value of $k_z$ where an extremal orbit exists and was observed by our quantum oscillatory experiments, i.e. the $\tau$ orbit in Fig.~\ref{fig:panel5}.
Notice that a gap of approximately $0.4 $ eV appears for this value of $k_z$. As noted in the Supplemental Information of Ref.~\onlinecite{kuan_wen}
an orbit situated slightly away from the DP can yield zero effective Berry-phase in quantum oscillation experiments.
Hence, this $\tau$ orbit is expected to yield a trivial Berry-phase in such experiments, as the distance with respect to the DP is considerable and leads to a large energy gap.
For reasons that remain unclear to us, and as seen in Fig.~\ref{fig:panel5}, the $\tau$ orbit is not detectable for fields oriented nearly along the \emph{a} or the \emph{c} crystallographic-axis.

The small ``islands'' at $\varepsilon_F$ in Fig.~\ref{fig:bands_kx_ky}(a), which correspond to the previously discussed small ellipsoid-like Fermi surface pockets seen in Fig.~\ref{fig:pdte2_bands}(b), exist only for a $k_z$ value in close proximity to $k_z^{\text{DP}}$ and have a maximum size at $k_z^{\text{DP}}$.
However, these ``islands'' emerge from a quadratic dispersion having a very distant connection to the Dirac type-II point.
Therefore, this orbit must also be topologically trivial.

\section{Summary}
To summarize,  we performed a detailed study on the quantum oscillatory phenomena observed in the Dirac type-II semimetallic candidates PdTe$_{2}$ and PtTe$_{2}$. We obtain very light effective masses, in the range of 0.04 to 1.5 electron mass for PdTe$_2$ and from 0.11 to 3.6 $m_0$ for PtTe$_2$, concomitant high mobilities in the order of $5 \times 10^{3}$ cm$^2$/Vs, and  a remarkable good agreement between density functional theory calculations and the topography of the Fermi surface of PdTe$_2$ as determined experimentally. Such agreement implies that this compound indeed displays a Dirac type-II node located at $\sim 0.5$ eV below the Fermi level. The agreement between the DFT calculations and the Fermi surface cross-sectional areas of PtTe$_2$ is relatively poor, although it does \emph{not} indicate a radical difference between the calculated and the experimentally determined Fermi surface topography. This suggests that this compound is also a good candidate for the existence of a Dirac type-II cone. However, in  PtTe$_2$ electronic band calculations indicate that the Dirac type-II cone would not intersect its Fermi level or that this compound would be characterized by topologically trivial charge carriers. Although the calculations indicate that the Dirac cone does intersect $\varepsilon_F$ in the case of PdTe$_2$, the associated orbit or Fermi surface cross-sectional area detected by quantum oscillations would be located on a different $k_z$ plane with respect to the plane of the Dirac type-II node. This small displacement is enough to lead to a Berry phase $\phi_{\text{B}} < \pi/2$ associated with topologically trivial \cite{kuan_wen} electronic orbits in PdTe$_2$. This is consistent with several recent reports claiming conventional superconductivity for PdTe$_2$. Nevertheless, it is still possible for PdTe$_2$ to display topologically non-trivial orbits at the Fermi level that would not coincide with the extremal cross-sectional areas detected by quantum oscillatory phenomena.

\section{Acknowledgment}
This work was supported by DOE-BES through award DE-SC0002613.
J.Y.C. acknowledges support from NSF through DMR-1700030.
We acknowledge the support of the HLD-HZDR, member of
the European Magnetic Field Laboratory (EMFL).
The NHMFL is supported by NSF through NSF-DMR-1157490 and the State of Florida.

\end{document}